\documentclass{elsart}

\usepackage{graphicx}
\usepackage{amssymb}
\usepackage{lineno}

\begin{document}
\begin{frontmatter}
\title{Distillation of Liquid Xenon to Remove Krypton}
\author{K. Abe\thanksref{label1}},
\author{J. Hosaka\thanksref{label1}},
\author{T. Iida\thanksref{label1}},
\author{M. Ikeda\thanksref{label1}},
\author{K. Kobayashi\thanksref{label1}},
\author{Y. Koshio\thanksref{label1}},
\author{A. Minamino\thanksref{label1}},
\author{M. Miura\thanksref{label1}},
\author{S. Moriyama\thanksref{label1}},
\author{M. Nakahata\thanksref{label1}\thanksref{label12}\corauthref{cor1}}, 
\ead{nakahata@suketto.icrr.u-tokyo.ac.jp}
\corauth[cor1]{Corresponding Author}
\author{Y. Nakajima\thanksref{label1}},
\author{T. Namba\thanksref{label1}\thanksref{preadd1}},
\author{H. Ogawa\thanksref{label1}},
\author{H. Sekiya\thanksref{label1}},
\author{M. Shiozawa\thanksref{label1}},
\author{Y. Suzuki\thanksref{label1}\thanksref{label12}},
\author{A. Takeda\thanksref{label1}},
\author{Y. Takeuchi\thanksref{label1}},
\author{K. Ueshima\thanksref{label1}},
\author{M. Yamashita\thanksref{label1}},
\author{K. Kaneyuki\thanksref{label11}},
% \author{T. Doke\thanksref{label2}},
\author{Y. Ebizuka\thanksref{label2}},
\author{J. Kikuchi\thanksref{label2}},
\author{A. Ota\thanksref{label2}},
\author{S. Suzuki\thanksref{label2}},
\author{T. Takahashi\thanksref{label2}},
\author{H. Hagiwara\thanksref{label3}},
\author{T. Kamei\thanksref{label3}},
\author{K. Miyamoto\thanksref{label3}},
\author{T. Nagase\thanksref{label3}},
\author{S. Nakamura\thanksref{label3}},
\author{Y. Ozaki\thanksref{label3}},
\author{T. Sato\thanksref{label3}},
\author{Y. Fukuda\thanksref{label4}},
\author{T. Sato\thanksref{label4}},
\author{K. Nishijima\thanksref{label5}},
\author{M. Sakurai\thanksref{label5}},
\author{T. Maruyama\thanksref{label5}},
\author{D. Motoki\thanksref{label5}},
\author{Y. Itow\thanksref{label6}},
\author{H. Ohsumi\thanksref{label7}},
\author{S. Tasaka\thanksref{label8}},
\author{S. B. Kim\thanksref{label9}},
\author{Y. D. Kim\thanksref{label10}},
\author{J. I. Lee\thanksref{label10}},
\author{S. H. Moon\thanksref{label10}},
\author{Y. Urakawa\thanksref{label15}},
\author{M. Uchino\thanksref{label15}},
and
\author{Y. Kamioka\thanksref{label15}}\\
for the XMASS Collaboration\\

\address[label1]{Kamioka Observatory, Institute for Cosmic Ray Research, The University of Tokyo, Kamioka, Hida, Gifu 506-1205, Japan}
\address[label11]{Research Center for Cosmic Neutrinos, Institute for Cosmic Ray Research, The University of Tokyo, Kashiwa, Chiba 277-8582, Japan}
\address[label12]{Institute for the Physics and Mathematics of the Universe, The University of Tokyo, Kashiwa, Chiba 277-8582, Japan}
\address[label2]{Faculty of Science and Engineering, Waseda University, Shinjuku-ku, Tokyo 162-8555, Japan }
\address[label3]{Department of Physics Faculty of Engineering, Yokohama National University, Yokohama, Kanagawa 240-8501, Japan}
\address[label4]{Department of Physics, Miyagi University of Education, Sendai, Miyagi 980-0845, Japan}
\address[label5]{Department of Physics, Tokai University, Hiratsuka, Kanagawa 259-1292, Japan}
\address[label6]{Solar Terrestrial Environment Laboratory, Nagoya University, Nagoya, Aichi 464-8602, Japan }
\address[label7]{Faculty of Culture and Education, Saga University, Honjo, Saga 840-8502, Japan}
\address[label8]{Department of Physics, Gifu University, Gifu, Gifu 501-1193, Japan}
\address[label9]{Department of Physics, Seoul National University, Seoul 151-742, Korea}
\address[label10]{Department of Physics, Sejong University, Seoul 143-747, Korea}
\thanks[preadd1]{Present address: International Centre for Elementary Particle
Physics, the University of Tokyo, Tokyo 113-0033, Japan}
\address[label15]{Taiyo Nippon Sanso Corporation, Tokyo Bldg. 1-3-26, Koyama,
Shinagawa-ku, Tokyo 142-8558, Japan}

\begin{abstract}
A high performance distillation system to remove krypton from xenon was
constructed, and a purity level of Kr/Xe = $\sim 3 \times 10^{-12}$ was
achieved. 
This development is crucial in facilitating high sensitivity low background
experiments such as the search for dark matter in the universe.
\end{abstract}

\end{frontmatter}

% in order to remove line number
% \linenumbers

\section{Introduction}
Liquid xenon is one of the most attractive materials for use in detectors for
astroparticle and particle physics \cite{bib:xmass, bib:xe10, bib:dama,
bib:zeplin, bib:meg}.
As a scintillator it has a large light yield, comparable to that of
NaI(Tl), and can be used for detectors with low energy
thresholds and high energy resolution. 
Because of the high atomic number of xenon ($Z = 54$) and its high density in 
liquid form ($\sim$ 3 g/cm$^3$), it contributes to the reduction of 
environmental backgrounds, such as $\gamma$-rays and
$\beta$-rays from uranium and thorium contamination, by self-shielding.
Another big advantage of liquid xenon is that xenon does not have long-lived
radioactive isotopes, and thus experiments on rare phenomena (such as
dark matter searches and double beta decay experiments) may be carried out
shortly after moving the xenon underground from the surface.

\begin{figure}
\begin{center}
\includegraphics[width=8cm]{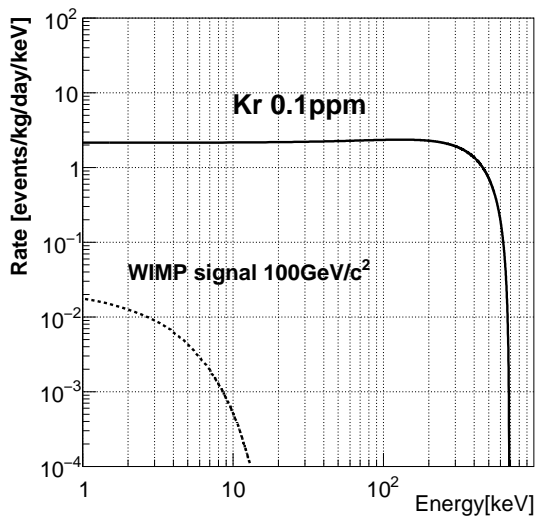}
\caption{Background event rate of $^{85}$Kr for a Kr/Xe ratio of 
$10^{-7}$ mol/mol, compared with dark matter signal rate, assuming
a cross section of $10^{-7}$ pb and a WIMP mass of 100 GeV/c$^2$. 
A quenching factor
of 0.2 is assumed for the dark matter signal.}
\label{fig:krbg}
\end{center}
\end{figure}

One drawback of liquid xenon is the possibility of contamination with 
radioactive $^{85}$Kr.
Xenon is usually produced from air. The concentration of xenon in air is 
$\sim 10^{-7}$ mol/mol, while the concentration of krypton is
$\sim 10^{-6}$ mol/mol.
%%%%%%%%%%%%%%%%%%%%%%%%%%%%%%%%%%%%%%%%%%%%%%%%%%%%%%%%%%%
In the commercial production of xenon, krypton is removed by 
distillation or adsorption. But, the commercially available xenon contains 
$10^{-9} \sim 10^{-6}$ mol/mol of krypton, because the removal starts
from the Kr/Xe ratio of about 10 and such purity is enough for most of the
applications of xenon.
%%%%%%%%%%%%%%%%%%%%%%%%%%%%%%%%%%%%%%%%%%%%%%%%%%%%%%%%%%%%
$^{85}$Kr is a radioactive nucleus which
decays into rubidium-85 with a half-life of 10.76 years, and emits
$\beta$-rays with a maximum energy of 687 keV and a 99.57\% branching ratio.
(The remaining 0.43\% represent a $\beta$-emission with a maximum energy of 
173 keV followed by a 514 keV $\gamma$-ray emission.)
Large quantities of $^{85}$Kr are produced artificially in nuclear fission.
This constitutes the main source of $^{85}$Kr in air.
The concentration of $^{85}$Kr in air is measured to be 
$\sim 1$ Bq/m$^3$\cite{bib:kr85a, bib:kr85b}, which corresponds to 
$^{85}$Kr/Kr = $\sim 10^{-11}$.
Assuming a Kr/Xe ratio of $10^{-7}$ mol/mol, the background event rate of
$^{85}$Kr is as shown in Fig. \ref{fig:krbg}.  
($^{85}$Kr/Kr of $1.2 \times 10^{-11}$ is used here.)
The expected dark matter event rate is also plotted, assuming
a cross section of $10^{-7}$ pb and a WIMP mass of 100 GeV/c$^2$. 
A quenching factor
of 0.2 is assumed for the dark matter signal.

Because next-generation dark matter detectors aim at a sensitivity of
10$^{-9}$ pb for the dark matter--proton cross section, 
a background rate of $10^{-4}$ events /keV/day/kg is required.
This corresponds to a Kr/Xe ratio of less than $\sim 10^{-12}$ 
mol/mol.

%%%%%%%%%%%%%%%%%%%%%%%%%%%%%%%%%%%%%%%%%%%%%%%%%%%%%%%%%%%%%%%%%%%%
Possible methods to remove krypton from xenon are distillation and 
adsorption.
They are commonly used industrial processes but systems which 
reduce Kr enough to meet our requirements did not exist.
A development of adsorption-based chromatography to achieve the required
Kr concentration is reported in ref.\cite{bib:chromato}.
%%%%%%%%%%%%%%%%%%%%%%%%%%%%%%%%%%%%%%%%%%%%%%%%%%%%%%%%%%%%%%%%%%%%
In this paper, we describe the development of a distillation system to
reduce krypton to a level of 10$^{-12}$.
In section 2, the design principle 
is discussed, while in section 3, the setup and operation of the distillation system 
are described. In section 4, we describe a technique for 
measuring low levels of krypton in xenon gas, 
and discuss the results.

\section{Design principle}

The boiling point of liquid krypton is 120 K at 1 atmosphere, while that of
xenon is 165 K. This means that, in principle, separation of krypton and xenon can be
performed by a distillation method. 
For the basic design of the distillation system, we followed the
McCabe-Thiele (M-T) method\cite{bib:mtmethod}.
%%%%%%%%%%%%%%%%%%%%%%%%%%%%%%%%%%%%%%%%%%%%%%%%%%%%%%%%%%%%%%%%%
It is the standard method of designing a distillation system.
At first, the concept of the M-T method is reviewed, and then the application
of the method for our purpose is described.
%%%%%%%%%%%%%%%%%%%%%%%%%%%%%%%%%%%%%%%%%%%%%%%%%%%%%%%%%%%%%%%%%

\begin{figure}
\begin{center}
\includegraphics[width=12cm]{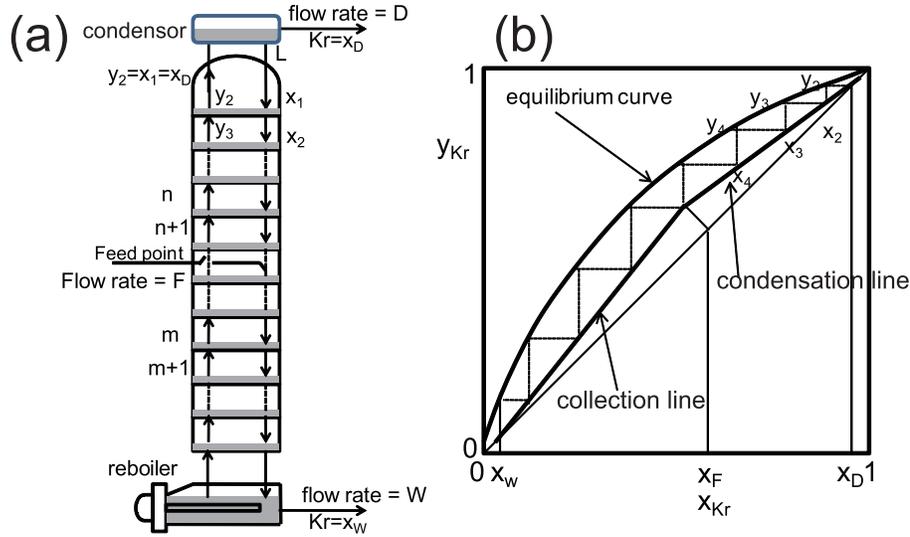}
\caption{(a) Illustration of the McCabe-Thiele (M-T) method.
The various elements of the figure are explained in the text.
(b) Calculation of the number of theoretical cells using the
M-T method. The horizontal and vertical axes represent the krypton concentration in the liquid 
and gas phases, respectively. The thick solid curve is the equilibrium 
curve, and the thick solid lines are the condensation line and the collection 
line (details explained in the text).}
\label{fig:design}
\end{center}
\end{figure}

The principle of the M-T method is shown in Fig. \ref{fig:design}.
The main element in the distillation system is a tower in which gas-liquid
equilibrium is maintained. A vessel called a ``reboiler'', at the bottom 
of the tower, boils the liquid using a heater. 
In order to maintain a constant temperature profile in the tower, 
a condenser is placed at its top.
The supplied xenon gas is cooled down to near boiling point, and then
supplied to the feed point in the tower at a flow rate $F$, as shown in
the figure.
The processed xenon, which contains a lower concentration of krypton than the
original xenon, is obtained from the reboiler (flow rate $W$), and
xenon with a higher krypton concentration is obtained at the top of the tower
(flow rate $D$). In the following, the Kr concentration of xenon at
these points is 
expressed as $x_F, x_W$ and $x_D$, respectively.
The heating power of the reboiler and the cooling power
of the condenser at the top of the tower control the flow of xenon 
in the tower (flow rate $L$).
The reflux ratio, represented by $R = L/D$, indicates
the amount of xenon returned to the tower compared with the amount extracted
with a high concentration of krypton.

In the M-T method, the distillation tower is assumed to be a connected series of
theoretical cells,
with the gas-liquid equilibrium changing by
one step in each cell.
The number of theoretical cells is calculated for given boundary conditions of
$F$, $D$, $W$, $x_F$, $x_W$, $x_D$ and $R$.

For an equilibrium of liquid and gas phases in a mixture of elements (in this case, Kr and Xe), 
the partial pressures of the gas-phase elements (expressed 
as $p_{Kr}$ and $p_{Xe}$, respectively) can be related to the fraction of 
each element 
in the liquid phase (expressed as $x_{Kr}$ and $x_{Xe}$) by Raoult's law:
\begin{eqnarray}
p_{Kr} &=& P_{Kr}\cdot x_{Kr}
\label{eq:Rao1}
\\
p_{Xe}&=& P_{Xe}\cdot x_{Xe}
\label{eq:Rao2}
\end{eqnarray}
where $P_{Kr}$ and $P_{Xe}$ are the vapor pressures of the corresponding elements 
(2090 kPa and 201.4 kPa, respectively, at 178 K).
Using Raoult's law, the concentrations of Kr in the gas phase ($y_{Kr}$)
and the liquid phase ($x_{Kr}$) can be related by the equation
\begin{eqnarray}
y_{Kr} = \frac{\alpha\cdot x_{Kr}}{1+(\alpha -1)x_{Kr}}
\label{eq:equi}
\end{eqnarray}
where $\alpha$ is the ratio of vapor pressures of Kr and Xe, $P_{Kr}/P_{Xe}$
(10.4 at 178 K).
Eq. \ref{eq:equi} is called the ``equilibrium curve'', and it is shown
by the thick solid curve in Fig. \ref{fig:design}(b).
For each theoretical cell, the gas- and liquid-phase Kr concentrations
in the neighboring cells are related by
conservation of mass flow, and may be expressed as:
\begin{eqnarray}
y_{n+1}&=& \frac{R}{R+1}x_{n}+\frac{1}{R+1}x_{d} \label{eq:cond} \\
y_{m+1}&=& \frac{R'}{R'-1}x_{m}-\frac{1}{R'-1}x_{W}
\label{eq:coll}
\end{eqnarray}
where eq. \ref{eq:cond} (\ref{eq:coll}) is for the cells above (below) the 
feeding point, and $n$ ($m$) is the number of theoretical cells.
$R'$ is $(L+qF)/W$, where $q$ is the fraction of liquid with respect to the
total feed material.
Equations \ref{eq:cond} and \ref{eq:coll} represent the ``condensation line''
and the ``collection line'', respectively; these are shown as solid lines in
Fig. \ref{fig:design}(b).
Figure \ref{fig:design}(b) illustrates the method of estimating the number of required theoretical cells.

\begin{figure}[hptb]
\begin{center}
\includegraphics[width=8cm]{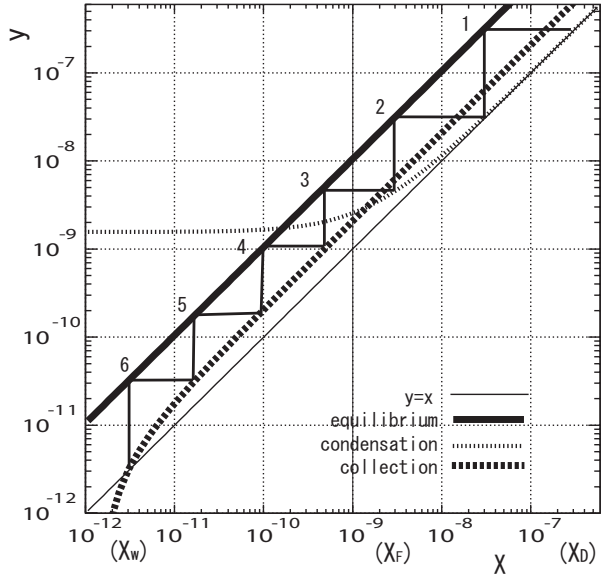}
\caption{M-T diagram with system requirements.}
\label{fig:diagram}
\end{center}
\end{figure}

We designed the distillation tower to fulfill the following requirements:
\begin{itemize}
\item The Kr concentration of the processed xenon should be three orders of magnitude smaller 
than that of the original xenon sample, i.e. $x_W = \frac{1}{1000} \times x_F$.
\item The collection efficiency of xenon should be 99\%, i.e. $W/F$ = 0.99 and 
$D/F$ = 0.01.
\item The system should have a process speed of 0.6 kg Xe per hour, which allows 100 kg xenon to be
purified within one week.
\item The system should have a reflux ratio of $R$ = 191, which means the heating power required
for the reboiler is 14 W.
\item Xenon should be fed into the system in the gas phase, i.e. $q$ = 0.
\end{itemize} 
A M-T diagram based on these requirements is shown in 
Fig. \ref{fig:diagram}, in which the concentration of Kr in original 
Xe($x_F$) was 
assumed to be $3 \times 10^{-9}$, the actual value of the processed xenon
described in section 4.
The M-T diagram shows that we need 6 stages of theoretical cells.

\section{Setup and operation of the distillation system}

A key element of the distillation system is the packing column in the 
distillation tower.
We used ``DX laboratory packing''(Sulzer Co.) with a diameter 
of 2 cm.
The catalog specifications of the packing
included a liquid load
(defined as the flow per unit area per unit time) 
between 0.1 and 5 $m^3/m^2h$.
In our case, the liquid load is 1.2 $m^3/m^2h$ for a reflux ratio $R$ of 191.
Gas loads, defined as $v \sqrt{\rho}$ (where $v$ and $\rho$ are gas
velocity and density, respectively), are 0.11 Pa$^{\frac{1}{2}}$ and 
0.23 Pa$^{\frac{1}{2}}$ for below and above the feed point, respectively.
The catalog of the packing gives the height equivalent to a theoretical plate
(HETP) for the distillation of a ethylbenzene/chlorobenzene mixture.
The given HETP value is 2$\sim$8 cm for the gas load of 
0.2-1.2 Pa$^{\frac{1}{2}}$.
However, since the HETP value might strongly depend on the liquid/gas loads 
and the type of liquid used, we conservatively enlarged the HETP value to
35 cm for the Kr/Xe distillation in the design, and the total length of the 
tower is set to 208 cm.
The feed enters the tower 10.8 cm from the top.

\begin{figure}[hptb]
\begin{center}
\includegraphics[width=12cm]{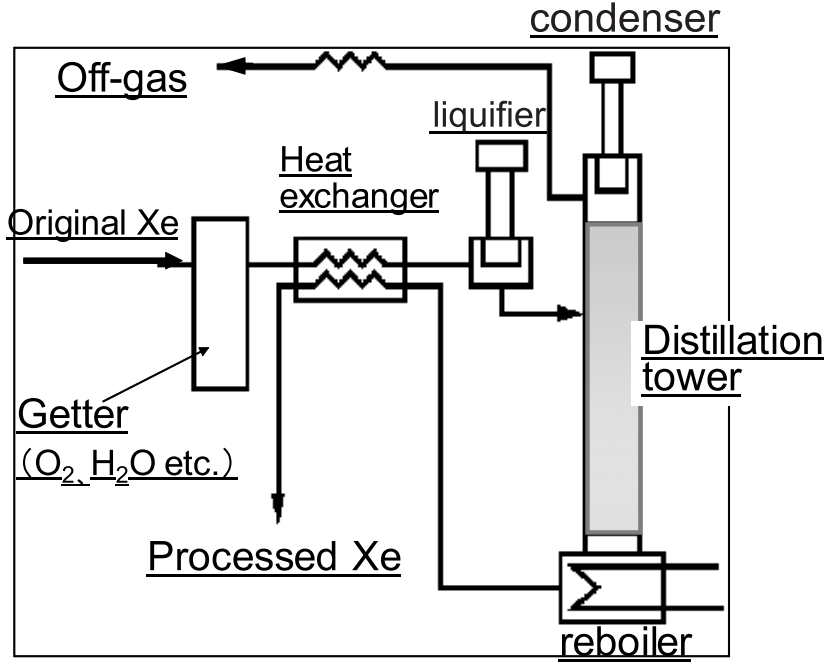}
\caption{Flow chart of distillation system.}
\label{fig:flow}
\end{center}
\end{figure}

The full setup of the distillation system is shown in Fig. \ref{fig:flow}.
Before entering the distillation apparatus, the xenon flows through a getter 
SPRG-100H-00030X (Taiyo Toyo Sanso Co.), which is able to decrease 
concentrations of
N$_2$, CH$_4$, O$_2$, CO$_2$, and CO to below $1 \times 10^{-10}$ mol/mol, and
H$_2$ and H$_2$O below $5 \times 10^{-10}$ mol/mol.
The xenon then flows through a heat exchanger which pre-cools the xenon.
For this purpose, the purified xenon gas extracted from the distillation 
apparatus is used as a cooling medium.
After the pre-cooling, the xenon is further cooled down to 192 K 
by a GM-type refrigerator 
(TZ-65; Taiyo Nippon Sanso Co.) which has a cooling power of 100 W 
at around 180 K, 
and the cooled xenon is fed into the distillation tower.
The temperature of the condenser at the top of the distillation tower is 
kept at 
178 K by an electrical heater.
The reboiler is a cylindrical copper vessel with a 200 mm diameter.
Three liquid level sensors are placed in the reboiler at 5 mm ($LL$
level), 20 mm ($L$ level) and 65 mm ($H$ level), corresponding to
0.47 kg, 1.88 kg and 6.1 kg of xenon, respectively.
The heater attached to the reboiler has a power of 22 W, and runs with a 65 \%
duty cycle during operation (this corresponds to 14 W effective power).
The purified xenon is collected from the reboiler
and goes to a collection bottle via the heat exchanger.
Xenon with a high Kr concentration, which is called the ``off-gas'', is collected 
from the top of the distillation tower. 
The purified and off-gas xenon were collected in stainless steel
bottles which were cooled by liquid nitrogen.

\begin{figure}[hptb]
\begin{center}
\includegraphics[width=8cm]{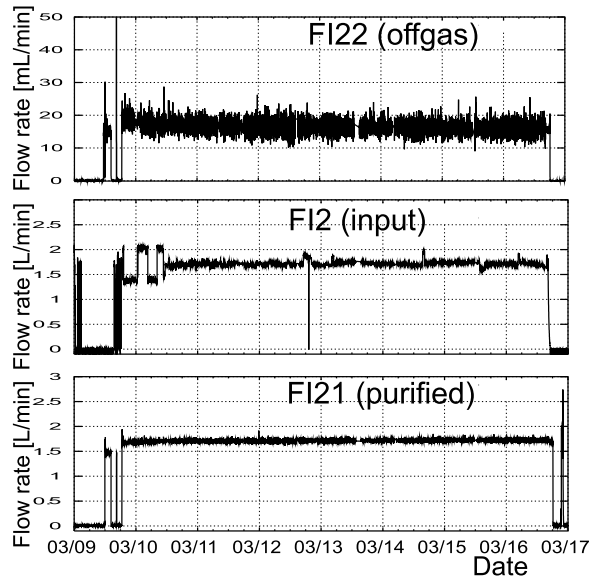}
\caption{Time variation of flow rate of off-gas xenon (upper figure), 
and input (middle) and purified xenon (lower).}
\label{fig:flowrate}
\end{center}
\end{figure}

\begin{figure}[hptb]
\begin{center}
\includegraphics[width=8cm]{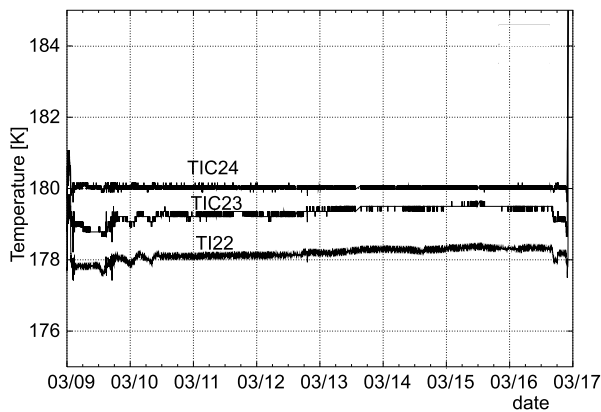}
\caption{Temperature variation of condenser,
(TI22), reboiler (TIC24) and center position in the tower (TIC23).}
\label{fig:temperature}
\end{center}
\end{figure}

The apparatus was used from March 9 to 17, 2004, to process 100 kg of xenon.
First, about 2.0 kg of xenon were liquefied and supplied to the 
distillation tower
up to liquid level $L$ in the reboiler.
The heater in the reboiler and the condenser at the top of the
distillation tower were switched on, and the system was maintained for 11 hours without 
collecting purified and off-gas xenon in order to establish
the initial gas-liquid equilibrium in the tower.
After this time, the flow rate of the purified xenon was set to 0.6 kg/h
(1.7 L/min gas flow), and that
of the off-gas xenon was set to 0.006 kg/h (17 mL/min).
The flow rate of the xenon feed was set at around 0.6 kg/h (1.7 L/min)
in the following manner: after the liquid level in the reboiler had reached $L$, the rate was set to 0.72 kg/h 
(2.0 L/min) for about four hours, 
and then to 
0.48 kg/h (1.4 L/min) until the xenon in the reboiler returned to level $L$.
When we found that the liquid level around $L$ could be measured
more precisely by the monitored pressure of the purified xenon, the flow
rate of the xenon feed was fine-tuned to balance the output flow.
Figure \ref{fig:flowrate} shows the time variation for off-gas, input and 
purified xenon.
Figure \ref{fig:temperature} shows the temperature variation as a function
of time for
the condenser in the distillation tower, the reboiler and the center 
position in the tower.
The temperature was stable to within $\pm$0.5 K during operation. 
The pressure in the distillation tower was approximately 0.105~MPa (gauge) 
throughout the operation. 

\section{Measurement of the krypton concentration}

In this section we describe the different methods used to measure
the krypton concentrations in the input xenon, off-gas, and 
purified xenon.

\subsection{Krypton concentration in the xenon feed}

\begin{figure}[hptb]
\begin{center}
\includegraphics[width=8cm]{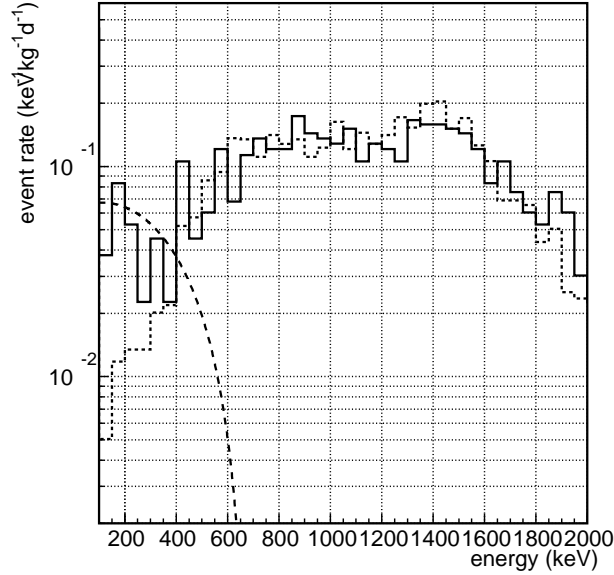}
\caption{The background spectra 
measured in December 2003 before the distillation(solid histogram) and in
August 2004 after the distillation(dotted histogram).
The dashed curve shows expected $^{85}$Kr beta spectrum assuming a Kr
concentration of 3$\times 10^{-9}$ mol/mol.}
\label{fig:xenonbg}
\end{center}
\end{figure}

The Kr concentration in the original xenon sample was evaluated by a
scintillation
measurement in a prototype detector for the XMASS experiment\cite{bib:xmass}.
The original xenon sample (100 kg) was placed in a 30 L chamber, 
and scintillation light was detected using
54 low-background photomultipliers\cite{prototype}. The background spectra 
were measured in December 2003 before the distillation\cite{noon2004} and in
August 2004 after the distillation\cite{ichep2004} as shown in 
Fig.\ref{fig:xenonbg}.
The background rate around 100--300 keV
before the distillation is higher than that after the distillation
by $\sim 5 \times 10^{-2}$ /kg/keV/day.
This background level corresponds to a Kr concentration of 
$(2 \sim 3) \times 10^{-9}$ mol/mol, assuming a
$^{85}$Kr/Kr ratio of 1.2$\times 10^{-11}$.

\subsection{Krypton concentration in the off-gas}

\begin{figure}[hptb]
\begin{center}
\includegraphics[width=10cm]{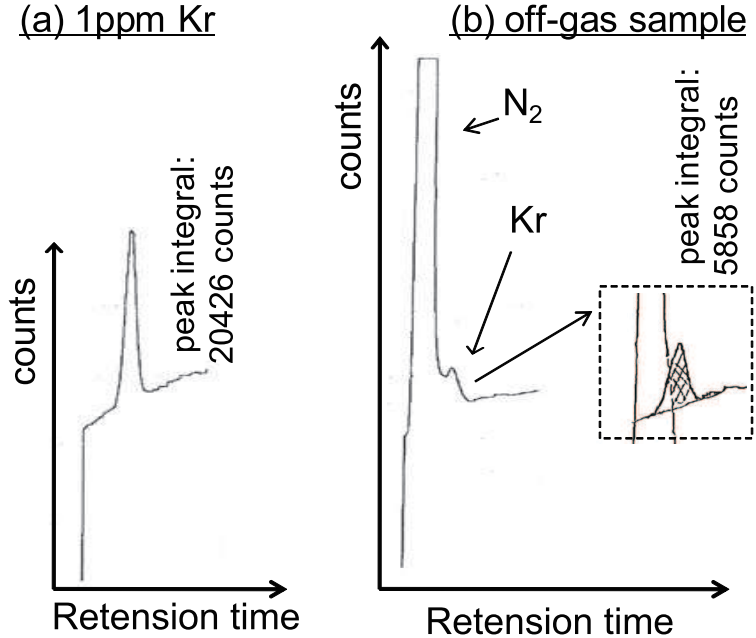}
\caption{Kr count by PID detector. The horizontal axis represents
travel time through the GC column. 
(a) Count rate for calibration
gas consisting of $1 \times 10^{-6}$ mol/mol Kr, 1\% Xe and 99\% helium. 
(b) Count rate for the off-gas sample.}
\label{fig:offgas}
\end{center}
\end{figure}

The Kr concentration of the off-gas xenon was measured on March 23, 2004.
Xe and Kr were separated by gas chromatography (GC), 
and the ion count was measured by a photoionization detector (PID).
The GC system for this experiment was a GC-263-50 (Hitachi Co.)
with a separation column consisting of a SUS tube of 3 mm inner diameter 
and 2 m length, which was filled with MS-13X molecular sieves (30/60 mesh).
In order to calibrate the system, a sample gas consisting of $1 \times 10^{-6}$
Kr, 1\% Xe and 99\% helium (1 ppm Kr gas) was measured using this setup,
and the count of Kr at the PID is shown in Fig. \ref{fig:offgas}(a) for a gas sample of 5 mL.
The integrated count rate for the Kr peak was 20,426.
The off-gas measurement is shown in Fig. \ref{fig:offgas}(b) for a sample of 5 mL.
The large peak in front of the Kr peak is due to the N$_2$ component in the
off-gas. As described in the previous section, the off-gas was collected by
a cold trap method using a collection bottle immersed in 
liquid nitrogen. Unfortunately, there was a small leak at the endcap of the
off-gas collection bottle.
The Kr concentration in the off-gas was evaluated by subtracting the
the tail of the N$_2$ contribution as shown in the figure, and the resulting
integral count rate of Kr in the off-gas was 5858.
Normalization of this value based on the calibration gas gave a Kr concentration in the
off-gas of $(3.3 \pm 1.0) \times 10^{-7}$ mol/mol.
The error estimate includes uncertainty introduced by the N$_2$ subtraction.

\subsection{Krypton concentration in the purified xenon}

A highly sensitive counting method was needed to measure the
Kr concentration in the purified Xe, because the concentration was expected to be
as low as $1 \times 10^{-12}$ mol/mol.
The measurement described below was carried out in September 2004.
We first describe the principle of the counting method.

The most sensitive method of gas component measurement which can be
applied to the measurement of Kr concentration is API-MS (atmospheric 
pressure ionization mass spectroscopy).
In this method, a sample gas can be introduced at near atmospheric 
pressure, which enables large amounts of gas to be loaded and consequently allows
highly sensitive measurements.
The gas that carries the sample into the API-MS system (the carrier gas)
must have a greater ionization energy than the target elements.
The carrier gas mixed with the sample gas is ionized at the first stage (at
this time, most of the ionized gas is carrier gas), and a
charge exchange interaction in the next stage transfers charge from the
carrier gas to the target elements. The ionized elements are then extracted 
and measured using a mass spectrometer.
We used helium as a carrier gas, because
the ionization energies of He and Kr are 24.6 eV and 14.0 eV, respectively.
Since the ionization energy of Xe (12.3 eV) is lower than that of Kr, 
Xe must be removed before introducing the sample into the API-MS. 
We used a GC technique for this purpose.
In order to load a large amount of the sample gas, we used a concentration
method, as detailed later.

\begin{figure}[hptb]
\begin{center}
\includegraphics[width=12cm]{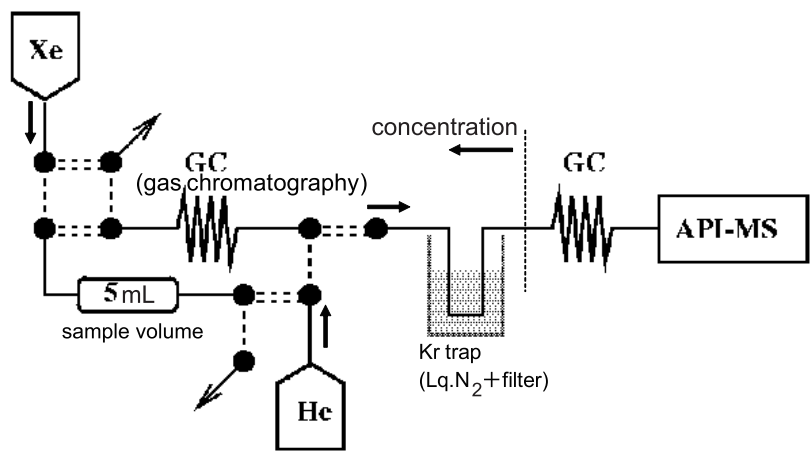}
\caption{Setup of high-sensitivity measurement of Kr in xenon gas.}
\label{fig:measetup}
\end{center}
\end{figure}
\begin{figure}[hptb]
\begin{center}
\includegraphics[width=8cm]{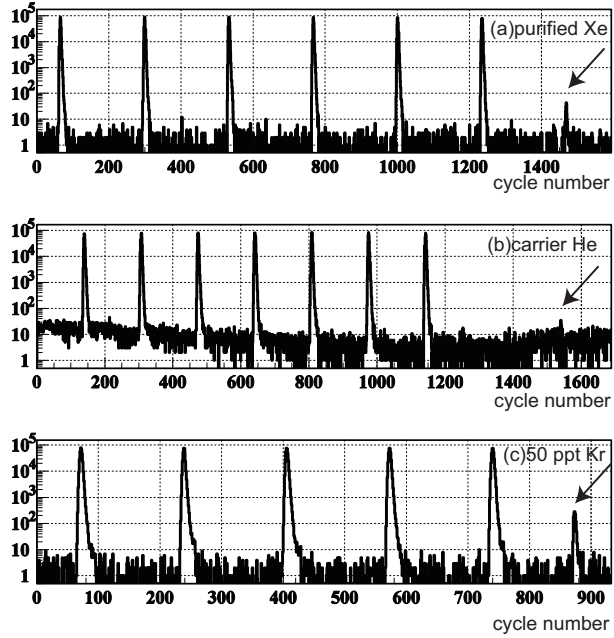}
\caption{Kr count rate measured by API-MS for (a) purified Xe gas,
(b) He carrier gas, and (c) 50 ppt Kr calibration gas.
Kr counting for each gas was conducted at the cycle marked by the arrow.
The high peaks appearing periodically before the real measurements represent
injections of 1 ppm Kr calibration gas, which is composed of $1\times 10^{-6}$ 
mol/mol Kr, 1\% Xe and 99\% He, and are used as a reference for overall
normalization.}
\label{fig:krcountall}
\end{center}
\end{figure}
\begin{figure}[hptb]
\begin{center}
\includegraphics[width=8cm]{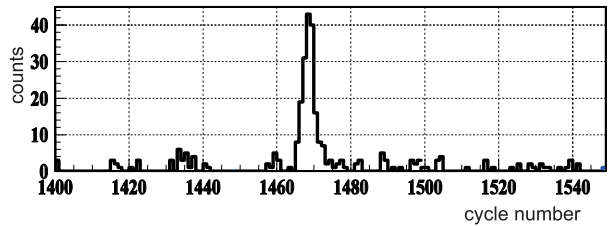}
\caption{Kr count rate in purified Xe. Expanded view of 
Fig. \ref{fig:krcountall} in the signal region.}
\label{fig:krcountxe}
\end{center}
\end{figure}

Figure \ref{fig:measetup} shows the setup for high-sensitivity Kr
measurement. 
First, the sample Xe gas was transferred to a 5 mL sample volume.
The 5 mL sample was then transferred to the GC system using
helium gas. 
The GC system used for this was GC-8APT (Shimadzu Co.) and
the separation column consisted of a SUS tube of 3 mm inner diameter 
and 2 m length, 
which was filled with MS-5A molecular sieves (30/60 mesh).
When the Kr arrived at the output line of the GC, the line was
connected to a metal filter (Nihon Seisen Co., pore size 0.003 $\mu$m) which
was cooled by liquid nitrogen, and the Kr was trapped by the metal filter.
The trapping efficiency was evaluated based on the results obtained for the calibration gas, and was
found to be as high as 80\%.
The processes described so far were repeated 100 times (concentration 
process), which enabled loading 500 mL (5 mL $\times$ 100 times) of the sample gas.
After concentration, the metal filter was warmed to room temperature by
removal of the liquid nitrogen vessel.
The helium carrier gas was then used to transfer the Kr trapped in the metal filter
to the API-MS. Another GC system was placed just in front of the API-MS to
separate Kr to ensure its counting.
The GC system used at this point had a separation column consisting of
a SUS tube of 3 mm inner diameter and 2 m length 
which was filled with MS-13X molecular sieves (30/60 mesh).
The API-MS apparatus was an API-200 (VG Gas Analysis Systems Ltd).
The count rate of the processed and concentrated Kr sample is shown in
Fig. \ref{fig:krcountall}(a).
The M/Z value (the mass number of the element divided by the charge) was set to 84 in this
measurement in order to count $^{84}$Kr$^{+}$ ions.
In order to check the stability of the API-MS system, a standard mixed gas 
composed of $1 \times 10^{-6}$ Kr, 1\% Xe and 99\% He (``1 ppm Kr'') 
was injected before the introduction of 
the sample gas.
The six peaks before cycle number 1300 in Fig.\ref{fig:krcountall}(a)
are due to the injection of the 
``1 ppm Kr''.
The real Kr signal of the purified Xe was counted at around cycle number 1470.
Figure~\ref{fig:krcountxe} shows a closer view of the region around the signal.
The signal count was defined as the integral counts within $_{-20}^{+30}$ cycles
of the peak, and was calculated to be 219 $\pm$ 14.8.
The background count of the API-MS was evaluated using the counts 
50 cycles before and 50 cycles after the signal region; the value obtained was 
0.70 $\pm$ 0.08 counts/cycle.
After subtracting the background contribution, the true Kr count of the 
purified Xe was
calculated to be 184.0 $\pm$ 15.8.
In order to estimate the Kr contamination in the carrier gas, the 
concentration process was repeated only using the He carrier gas (i.e.
without introducing the 5 mL xenon sample during each cycle). 
The API-MS count for the He carrier gas is shown in 
Fig. \ref{fig:krcountall}(b).
Similarly, in order to evaluate the overall counting efficiency, which incorporates
the collection efficiency of Kr and the counting efficiency of the API-MS system, a 
calibration gas containing $5 \times 10^{-11}$ mol/mol Kr
(50 ppt Kr) was processed
using the same method; its count rate is shown in 
Fig. \ref{fig:krcountall}(c).
The initial signal counts, background count rates and true
signal counts (with background rate subtracted) for the purified Xe sample, the He carrier gas and 
the 50 ppt Kr gas are summarized in Table~\ref{tbl:krcount}.
The true signal is corrected for the variation of their respective 
``1 ppm Kr'' signals.
Arbitrarily setting this normalization to 1.0 for the purified Xe, 
the corrections for the He carrier gas and the 50 ppt Kr become 0.92 and 0.88 
respectively. 

\begin{table}[h]
 \begin{center}
  \begin{tabular}[H]{|c|c|c|c|}
   \hline
   -& Purified Xe & He carrier gas & 50 ppt Kr \\
   \hline
   Signal & 219$\pm$14.8 & 409$\pm$22.2 & 1235$\pm$35.1 \\
   \hline
   BG~count/cycle & 0.70$\pm$0.08 & 6.41$\pm$0.25 & 1.16$\pm$0.12 \\
   \hline
   True signal(S) & 184.0$\pm$15.8 & 95.6$\pm$25.8 & 1341.7$\pm$40.7 \\
   \hline
  \end{tabular}
  \end{center}
  \caption{Signal counts, background count rates and 
true signal counts for purified Xe, He carrier gas, and 50 ppt Kr gas.
The true signal is $(signal - BG/cycle \times 50 cycles)/factor$, where the
$factor$ is the relative counting rate of the ``1ppm Kr''.}
 \label{tbl:krcount}
\end{table}

The Kr concentration of the purified Xe was calculated using the
following formula:
\begin{eqnarray}
\mathrm{Kr\ concentration} = \{(\mathrm{Purified\ Xe})-
(\mathrm{He\ carrier\ gas})\}\times
\frac{5\times 10^{-11}}{(50\ \mathrm{ppt\ Kr})}
\nonumber
\end{eqnarray}
where the values in round parentheses are the true signal counts given in Table
\ref{tbl:krcount}.
The value obtained for Kr concentration was
$(3.3 \pm 1.1) \times 10^{-12} ~~ \mathrm{Kr/Xe\ [mol/mol]}$.

\section{Conclusion}

A distillation system for removing Kr from Xe down to a concentration of
$\sim 10^{-12}$ Kr/Xe[mol/mol] was developed.
The system was designed to reduce Kr concentration by three orders of 
magnitude with 99\% Xe collection efficiency (i.e., the amount of Xe rejected is
only 1\%) and with a process speed of 0.6 kg Xe/h. 
This distillation system was used to purify 100 kg xenon gas containing
$3 \times 10^{-9}$ Kr mol/mol.
The off-gas was found to have a Kr concentration of $(3.3 \pm 1.0) \times 10^{-7}$ mol/mol,
which is consistent with the transfer of the majority of Kr in the original Xe sample
to the off-gas.
%In order to measure the Kr concentration of the purified Xe, a highly 
%sensitive
%measurement method was developed using a gas chromatography apparatus, a metal %filter to
%trap Kr, and a API mass spectrometer.
The Kr concentration of the purified Xe was measured as 
$(3.3 \pm 1.1) \times 10^{-12}$ Kr/Xe[mol/mol]
%%%%%%%%%%%%%%%%%%%%%%%%%%%%%%%%%%%%%%%%%%%%%%%%%%%%%%%%%%%%%%%%%%%%%%%%
using a gas chromatography apparatus and a API mass spectrometer.
%%%%%%%%%%%%%%%%%%%%%%%%%%%%%%%%%%%%%%%%%%%%%%%%%%%%%%%%%%%%%%%%%%%%%%%%
These measurements confirmed that the distillation system can
reduce the amount of Kr in xenon gas by up to three orders of magnitude.
The achieved Kr concentration satisfy our requirements for the next
generation of dark matter experiments.
 
\section{Acknowledgements}
 We gratefully acknowledge the cooperation of Kamioka Mining and Smelting Company.
 This work was supported by the Japanese Ministry of Education, Culture, Sports, Science and Technology, and a Grant-in-Aid for Scientific Research.

\end{document}